\documentclass[12pt]{iopart}

\usepackage{iopams}
\usepackage{setstack}
\usepackage{xspace}
\usepackage{graphicx}
\usepackage{hyperref}
\usepackage{lineno}

\newcommand{\pt}{\ensuremath{p_{\rm T}}\xspace}

\newcommand{\rt}{\ensuremath{R_{\rm T}}\xspace}

\newcommand{\nch}{\ensuremath{N_{\rm ch}}\xspace}

\newcommand{\nmpi}{\ensuremath{N_{\rm mpi}}\xspace}

\newcommand{\py}{PYTHIA\xspace}
\newcommand{\pPb}{p--Pb\xspace}
\newcommand{\pp}{pp\xspace}

\begin{document}


\title[]{Investigating the most active pp collisions (top 0.1\%) using the tools developed by experiments at the LHC}

\author{Jes\'us Eduardo Mu\~noz M\'endez and Antonio Ortiz}
\ead{jesus.munoz@correo.nucleares.unam.mx}

\address{Instituto de Ciencias Nucleares, UNAM, Apartado Postal 70-543, Coyoac\'an, 04510, Mexico City, Mexico}

\vspace{10pt}
\begin{indented}
\item[]\today
\end{indented}

\begin{abstract}
The LHC data have unveiled unexpected features in proton-proton (pp) collisions, namely, collective-like behavior and strangeness enhancement. Originally, these new effects were discovered only in high-multiplicity pp collisions. However, recently the ALICE Collaboration has shown that even low-multiplicity pp collisions yield a non-zero elliptic flow ($v_{2}$). Moreover, analyses as functions of the event structure, such as transverse spherocity, suggest that multiplicity might not be the main driver of the new effects. Therefore, new ways of analyzing the data have to be explored in order to understand the origin of the new phenomena. In this paper, pp collisions simulated with \py~8 are analyzed using different event estimators (mid-pseudorapidity multiplicity, spherocity, sphericity, $R_{\rm T}$, forward multiplicity and flattenicity). The features of the selected events for the top 0.1\% using the different estimators are discussed. The transverse momentum spectrum of primary particles and the recoil jet distributions are analyzed. The results suggest that flattenicity is the estimator with the least bias on the neutral-to-charged particle ratio, and on the bias towards harder-than-average pp collisions.   
\end{abstract}

%
\noindent{\it Keywords}: LHC, flattenicity, QCD \\
%
%
%
%

\section{Introduction}\label{sec:1}

One of the main results of the last decade is the discovery of heavy-ion-like behavior in small collision systems: \pp and \pPb interactions. However, the origin of these new findings remains an open question. In particular because in heavy-ion collisions there is strong evidence that supports the formation of the strongly-interacting quark--gluon plasma (sQGP) in central collisions~\cite{BRAHMS:2004adc,PHENIX:2004vcz,PHOBOS:2004zne,STAR:2005gfr,ALICE:2022wpn}, and there, jet quenching effects have been observed which are absent in small collision systems. 

The studies of high-multiplicity pp collisions have been conducted by experiments at the RHIC and LHC. The CMS Collaboration discovered the ridge structure in events with an average track multiplicity of 136 considering charged particles in $|\eta|<2.4$ and $p_{\rm T}>$0.4\,GeV/$c$~\cite{CMS:2010ifv}. Meanwhile, the ALICE Collaboration discovered deviations, with respect to QCD-inspired event generators, of the average transverse sphericity with increasing multiplicity of charged particles in pp collisions at $\sqrt{s}=7$\,TeV~\cite{ALICE:2012cor}. Later, ALICE found strangeness enhancement in pp collisions selecting events based on the amplitude measured in the forward V0 detector~\cite{ALICE:2016fzo}. 

Although hydrodynamics has been applied to explain the \pp and \pPb data, some theoretical works suggest that it should not be applicable in small collision systems~\cite{Ambrus:2023qvk}. Based on IP-Glasma calculations~\cite{Dumitru:2008wn,Bzdak:2013zma}, the energy density in high-multiplicity pp collisions, averaged over the transverse area, can be around 70\,GeV/fm$^{3}$ for charged particle multiplicity densities above 100~\cite{ALICE:2020fuk}. From lattice QCD calculations, this condition would be enough to expect the formation of the strongly interacting quark--gluon plasma in pp collisions. However, no signatures of jet quenching have been observed so far~\cite{ALICE:2023plt}. This opens the door to alternative approaches, such as color strings, that have successfully described some aspects of the data~\cite{Bierlich:2014xba}. A recent implementation of heavy-ion collisions in \py (Angantyr)~\cite{Bierlich:2018xfw} opened up a new venue to explore dense QCD systems. In particular, \py/Angantyr is able to describe the low transverse momentum yield ($p_{\rm T}<3$\,GeV/$c$) in central heavy-ion collisions within 20\%. This is a remarkable achievement, since the low \pt region is the most sensitive to the bulk properties of sQGP. Given that \py does a relatively good job describing many aspects of the LHC data, the present paper uses \py version 8.312 tune Monash~\cite{Skands:2014pea} (\py~8 from now on) to explore new ways of analyzing the \pp data.

The most natural way to select high-multiplicity pp collisions is by imposing a threshold in the number of primary charged particles at mid-pseudorapidity. However, since in general the multiplicity is expected to bias the event sample towards harder-than-average pp collisions, other event activity classifiers were proposed. In the context of the present study, the event activity refers to the collection of particles produced by multi-parton interactions~\cite{Sjostrand:1987su}. Our work presents a study of the different event classifiers that have been studied by experiments so far, and the features of the selected samples are compared with each other. The top 0.1\% of each event classifier distribution, containing the pp collisions with the highest event activity, are analyzed.  The 0.1\% fraction is obtained cutting at different values of each event classifier: multiplicity of charged particles in both mid-pseudorapidity ($N_{\rm ch}$) and forward pseudorapidity (V0M), as well as transverse sphericity ($S_{\rm T}$) ~\cite{Ortiz:2011pu,ALICE:2012cor,CMS:2025sws}, spherocity ($S_{0}$)~\cite{Cuautle:2015kra,Acharya:2019mzb}, the relative transverse activity classifier ($R_{\rm T}$)~\cite{Martin:2016igp,ALICE:2023csm,ALICE:2023yuk}, and flattenicity ($\rho_{\rm nch}$)~\cite{Ortiz:2022mfv, ALICE:2024vaf}. Since the largest multiplicity reach is only attainable using the mid-pseudorapidity multiplicity estimator, for this event classifier the largest energy density is achieved. The studies presented here are relevant for helping interpret the plethora of data which have been published by experiments at the LHC and understand whether a compatibility between the different event classifiers exists.

The paper is organized as follows. Section~\ref{sec:2} introduces the event activity classifiers that are considered in this paper. The biases on the charged-to-neutral particle yields, as well as on the selection of harder-than-average pp collisions, are discussed in Sec.~\ref{sec:3} and Sec.~\ref{sec:4}, respectively. The correlations among all the event activity estimators, as well as a discussion on the prospects for the data analysis of the LHC Runs 3 and 4, are presented in Sec.~\ref{sec:5}. A summary of the main findings is presented in Sec.~\ref{sec:6}. 

\section{Event classifiers}\label{sec:2}

In the present work, 6 billion \pp collisions at $\sqrt{s}=13$\,TeV (equivalent to an integrated luminosity of around 70\,nb$^{-1}$) are simulated with \py~8, the simulations do not include pile-up effects. The Monash \py tune is used, as it describes many observables involving unidentified charged particles measured at the LHC. The tune uses the leading-order parton distribution function NNPDF2.3~\cite{Ball:2013hta}. The event classifiers listed below are calculated using primary-charged particles, and for pp collisions with at least one particle ($p_{\rm T}>0$) within $|\eta|<1$. The collection of pp collisions satisfying this event selection criterion is named the minimum-bias (MB) sample. This list is not exhaustive, as other event classifiers used in the community are not included in the present study. The primary-charged particles are defined as particles with a mean proper lifetime greater than 1\,cm$/c$, which are produced directly in the interaction or from decays of particles with a mean proper lifetime smaller than 1\,cm$/c$~\cite{ALICE:2017hcy}.

{\bf Mid-pseudorapidity multiplicity estimator:} this is the most basic event classifier that is defined as the number of primary-charged particles in the pseudo-rapidity interval of $-0.8<\eta<0.8$, corresponding to the midrapidity region considered in ALICE studies.

{\bf V0M multiplicity estimator:} this event classifier considers the particle multiplicity registered in the forward pseudorapidity region that is available in the ALICE detector. Only primary-charged particles that enter into any of the sectors covered by the V0 detector are considered. Table~\ref{tab:1} displays the segmentation of the V0 detector that corresponds to the selection considered in the present study.

\begin{table}
\centering
                \caption{Pseudorapidity intervals covered by the different rings of the V0 detector of ALICE.
                 \label{tab:1}}
                 \begin{indented}
                \item[]\begin{tabular}{ccc}
                \br
                \textbf{Ring}  & \textbf{V0C} & \textbf{V0A} \\ 
                \mr
                1       & $-3.7<\eta<-3.2$  & $4.5<\eta<5.1$  \\
                2       & $-3.2<\eta<-2.7$  & $3.9<\eta<4.5$  \\
                3       & $-2.7<\eta<-2.2$  &  $3.4<\eta<3.9$ \\
                4       & $-2.2<\eta<-1.7$  &  $2.8<\eta<3.4$ \\
                \br
                \end{tabular}
                \end{indented}
\end{table}

{\bf Transverse sphericity estimator:} in hadron colliders, event shapes are defined in the transverse plane, i.e. in the plane perpendicular to the beam axis. Only primary-charged particles at mid-pseudorapidity ($|\eta|$$<$0.8) and $p_{\rm T}>0$ are considered.  This particle-level selection criterion is accessible to experiments such as ALICE~\cite{Abelev:2012sk}, CMS~\cite{CMS:2011usu} and ATLAS~\cite{ATLAS:2012uka}.  For the transverse sphericity to be defined, the transverse momentum matrix, $\bf{S}$, should be first diagonalized:

\begin{equation}
\mathbf{S} =  \frac{1}{\sum_{i} p_{{\rm T},i}}  \sum_{i} \frac{1}{p_{{\rm T},i}}
\left( {\begin{array}{*{20}c}
   p_{{\rm x},i}^{2} &     p_{{\rm x},i}p_{{\rm y},i}  \\
  p_{{\rm y},i}p_{{\rm x},i}  &   p_{{\rm y},i}^{2} \\
 \end{array} } \right)
\end{equation}

where, $p_{{\rm T},i}$ is the transverse momentum of the $i$-th particle, being $p_{{\rm x},i}$ and $p_{{\rm y},i}$ the components along the x and y axes, respectively. The transverse sphericity is then defined in terms of the eigenvalues, $\lambda_{1}$$>$$\lambda_{2}$, as follows:

\begin{equation}
S_{\rm T} \equiv \frac{2\lambda_{2}}{\lambda_{1}+\lambda_{2}}.
\end{equation}

{\bf Transverse spherocity:}  originally proposed here~\cite{Banfi:2010xy} and studied in~\cite{Cuautle:2015kra} for soft QCD physics is defined for a unit vector ${\hat{\rm \mathbf{n}}_{\rm \mathbf{s}} }$ which minimizes the ratio:

\begin{equation}
S_{\rm 0} \equiv \frac{\pi^{2}}{4}  \underset{\bf \hat{n}_{\rm \bf{s}}}{\rm{min}}  \left( \frac{\sum_{i}|{\vec p}_{{\rm T},i} \times { \bf \hat{n}_{\rm \bf{s}} }|}{\sum_{i}p_{{\rm T},i}}  \right)^{2}.
\end{equation}

Spherocity is calculated considering primary-charged particles at mid-pseudorapidity ($|\eta|$$<$0.8) and $p_{\rm T}>0$.

By construction, the limits of transverse sphericity and spherocity are related to specific configurations in the transverse plane~\cite{Ayala:2009jw}:

\begin{equation}
S_{\rm T},S_{0} \rightarrow 
\cases{0 & \rm{``pencil-like'' limit}\\
      1 & \rm{``isotropic'' limit}\\}
\end{equation}

{\bf Relative-transverse activity classifier:} as one selects high multiplicities in small collision systems, the event sample is naturally biased towards hard processes. To overcome such selection biases, the jet signal can be explicitly removed from the multiplicity estimator. This is done by building the azimuthal correlations of associated particles with the leading one. The leading particle is the charged particle with the largest transverse momentum in the event ($p_{\rm T}^{\rm lead}$), the associated particles satisfy $p_{\rm T}<p_{\rm T}^{\rm lead}$. The particles are selected within the pseudorapidity interval $-0.8<\eta<0.8$ and $p_{\rm T}>0$. The leading particle should have a transverse momentum within 5-40\,GeV/$c$, since in this region the average underlying-event activity saturates. The transverse region is defined relative to the leading particle; this is formed by particles within $\pi/3<|\Delta\varphi|<2\pi/3$, where $\Delta\varphi=\varphi-\varphi^{\rm lead}$ and $\varphi^{\rm lead}$ ($\varphi$) is the azimuthal angle of the leading (associate) particle. The transverse region is used to build a new event classifier \rt that has a low sensitivity to hard processes~\cite{Martin:2016igp}. The relative transverse activity classifier, \rt, is defined as the ratio of the multiplicity of the primary charged particles in the transverse region ($N_{\mathrm{ch}}^{\mathrm{T}}$) obtained event-by-event to the average value ($\langle N_{\mathrm{ch}}^{\mathrm{T}} \rangle$) ~\cite{Martin:2016igp,ALICE:2019mmy}. It is given as
    
    \begin{equation}\rt=\frac{N^{\mathrm{T}}_{\mathrm{ch}}}{\langle N^{\mathrm{T}}_{\mathrm{ch}} \rangle}.
        \label{Rt}
    \end{equation}
    
The pp collisions with $\rt \rightarrow0$ are associated with clean dijet structures since by construction little or no underlying event will be present. On the other hand, the pp collisions with large \rt values will be those in which the paticle density from underlying event activity as well as from initial- and final-state radiation will be significantly higher than that for the jet-like yield~\cite{Bencedi:2020qpi}.  
    
{\bf Flattenicity:} inspired by flatenicity~\cite{Ortiz:2022zqr}, which is proposed as a new observable to be measured in the next generation heavy-ion experiment at CERN (ALICE 3) in the LHC Run~5~\cite{ALICE:2803563}. The flattenicity was redefined to classify events using the existing detectors of ALICE~\cite{ALICE:2014sbx},

\begin{equation}
\rho_{\rm nch}=\frac{\sqrt{\sum_{i}{(N_{\rm ch}^{{\rm cell},i} - \langle N_{\rm ch}^{\rm cell} \rangle)^{2}}/{\rm N_{cell}^{2}}}}{\langle N_{\rm ch}^{\rm cell} \rangle},
\end{equation} 
where, $N_{\rm ch}^{{\rm cell},i}$ is the average multiplicity in the elementary cell $i$ and $\langle N_{\rm ch}^{\rm cell} \rangle$ is the average of $N_{\rm ch}^{{\rm cell},i}$ in the event. Flattenicity is calculated in the pseudorapidity intervals covered by the ALICE V0 detector (see Tab.~\ref{tab:1}) and using primary charged particles with $p_{\rm T}>0$. The additional factor $1/{\rm N_{cell}}$ guarantees flattenicity to be smaller than unity. Moreover, in order to have a similar meaning of the limits of the new event shape to those used so far (e.g. spherocity), this paper reports results as a function of $1-\rho_{\rm nch}$, in such a way that events with $1-\rho_{\rm nch}\rightarrow1$ are associated with the ``isotropic'' topology (in the azimuthal angle and pseudorapidity), whereas those with $1-\rho_{\rm nch}\rightarrow0$ are associated with ``pencil-like'' topologies.

\begin{figure}[!h]
\centering
\includegraphics[width=0.32\textwidth]{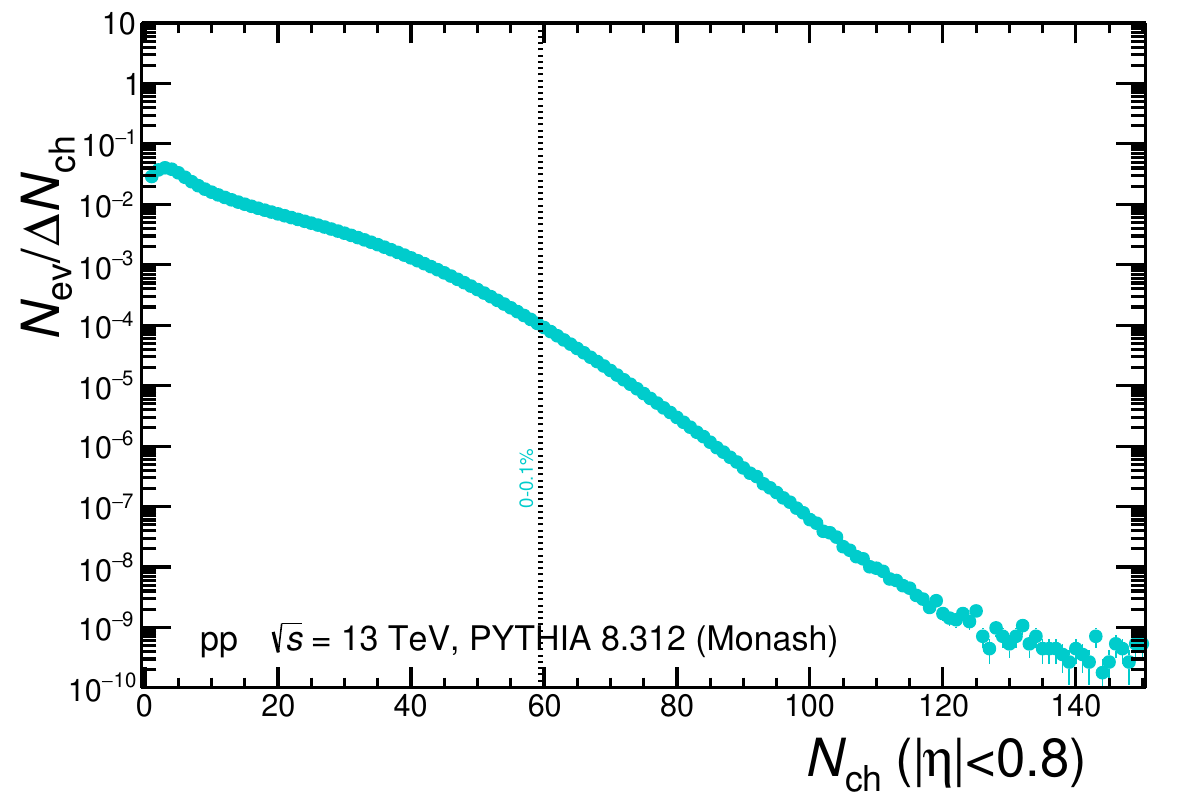}
\includegraphics[width=0.32\textwidth]{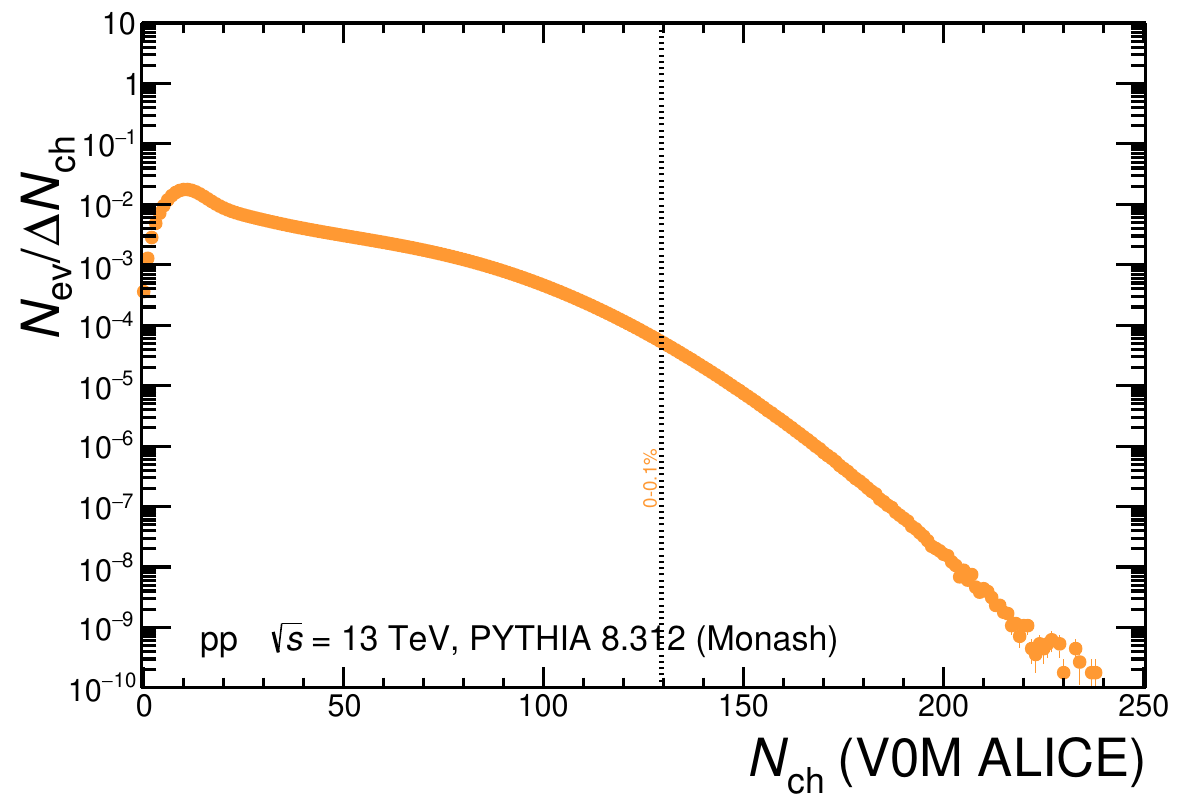}
\includegraphics[width=0.32\textwidth]{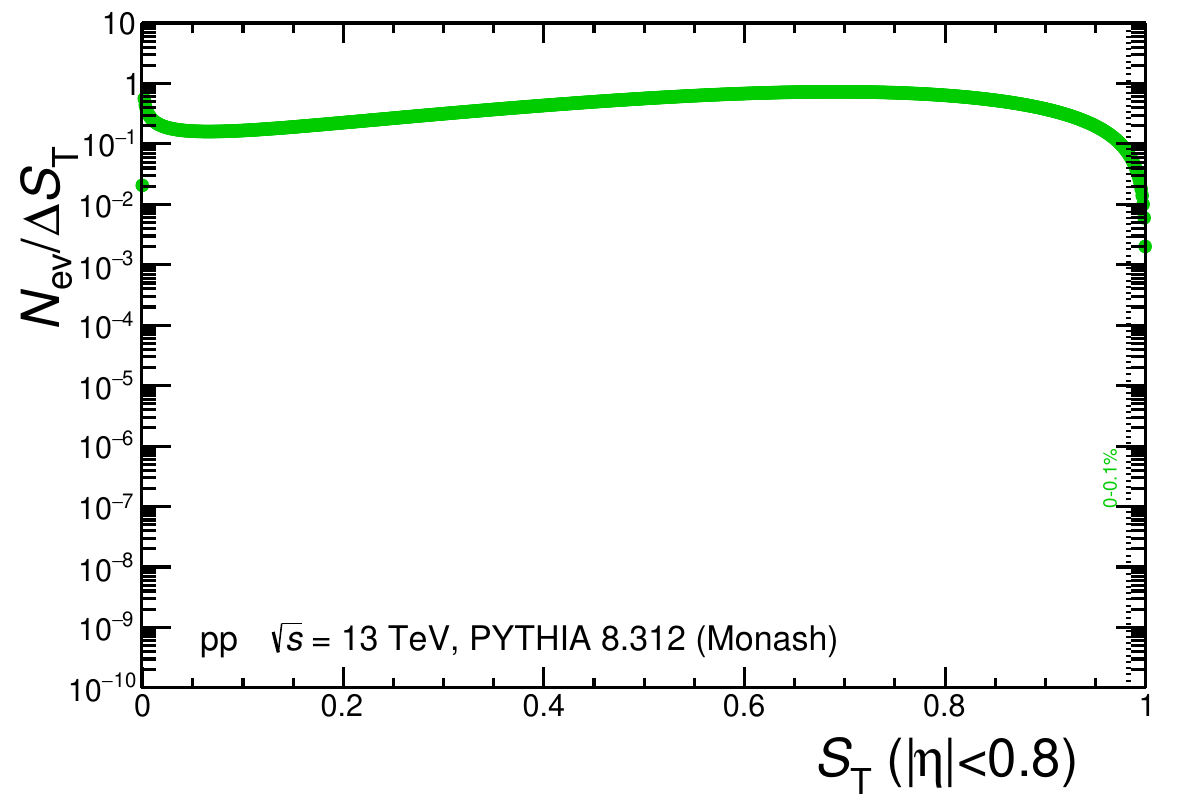}
\includegraphics[width=0.32\textwidth]{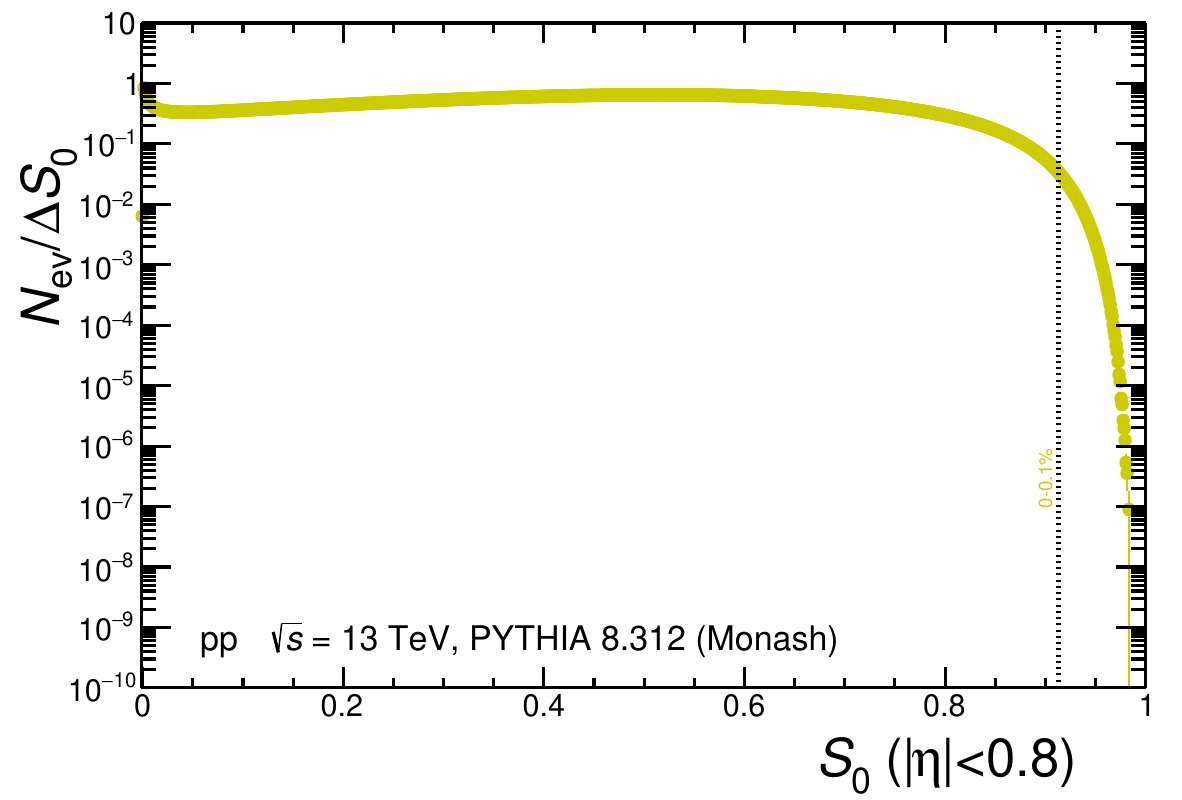}
\includegraphics[width=0.32\textwidth]{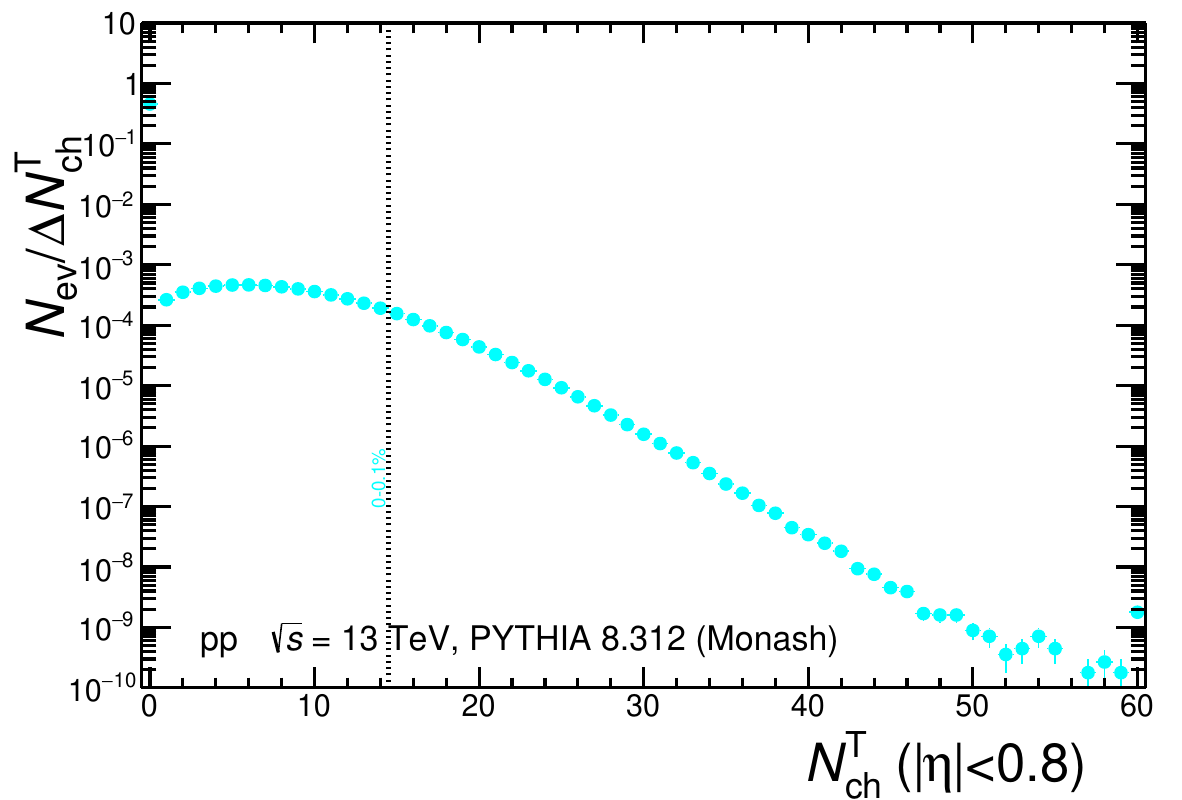}
\includegraphics[width=0.32\textwidth]{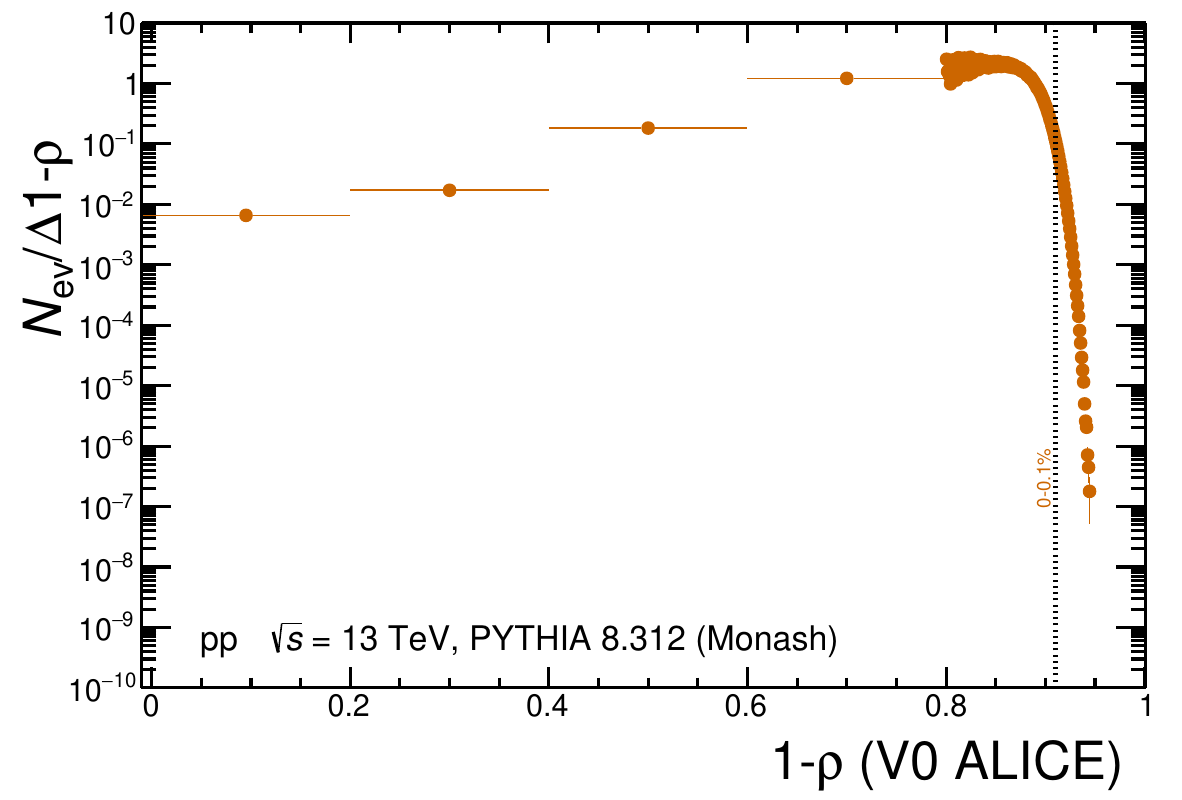}
\caption{Distribitions of the event activity estimators calculated in pp collisions at $\sqrt{s}=13$\,TeV with \py~8. The error bars represent the statistical uncertainty. The dashed lines indicate the top 0.1\% event classes.}
\label{fig:0}  
\end{figure}

The characteristics of the estimators are shown in Table~\ref{tab:2}. Although V0M and flattenicity are calculated in the forward pseudorapidity regions, flattenicity uses information both from pseudorapidity and azimuthal angle. In contrast, the other estimators ($N_{\rm ch}$, \rt, $S_{T}$ and $S_{0}$) are calculated at mid-pseudorapidity. Figure~\ref{fig:0} shows the distributions of all the event activity estimators discussed above. The calculation was done for the 6 billion minimum-bias pp collisions and for events that satisfy the event selection (at least one primary-charged particle $p_{\rm T}>0$ within $|\eta|<1$). In all the cases, the maximum values of the estimators are associated with pp collisions with large event activity.  

\begin{table}[!h]
\scriptsize
                \caption{The different event classifiers explored by experiments at the LHC, in particular by the ALICE Collaboration. The pseudorapidity intervals considered for the calculation of the classifiers are specified in the table.
                 \label{tab:2}}
                \begin{tabular}{@{}lccccc} 
                \br
                \textbf{Event classifier}  &   \textbf{Symbol} & \textbf{$\mathbf{\eta}$ coverage}  & \textbf{$\varphi$ coverage (rad)} &  \textbf{is $\eta$ used ?} & \textbf{is $\varphi$ used ?} \\ 
                \mr
                 Multiplicity  &  $N_{\rm ch}$   & $-0.8<\eta<0.8$  & $2\pi$ & yes & no \\
                 Sphericity      &  $S_{\rm T}$ &   $-0.8<\eta<0.8$  & $2\pi$ & yes & no \\
                 Spherocity      &  $S_{0}$ &   $-0.8<\eta<0.8$  & $2\pi$ & yes & no \\                 
                 Rel. trans. act. class.  &  $R_{\rm T}$   & $-0.8<\eta<0.8$  &  $\pi/3<|\Delta\varphi|<2\pi/3$ & yes & yes \\
                 Forward multiplicity     & V0M   & $\eta$ covered by V0 (see Table~\ref{tab:1})  & $2\pi$ & yes & no \\
                 Flattenicity      &   1-$\rho_{\rm nch}$ & $\eta$ covered by V0 (see Table~\ref{tab:1})  & $2\pi$ & yes & yes \\
                \br
                \end{tabular}
\end{table}

\begin{figure}[!h]
\centering
\includegraphics[width=0.6\textwidth]{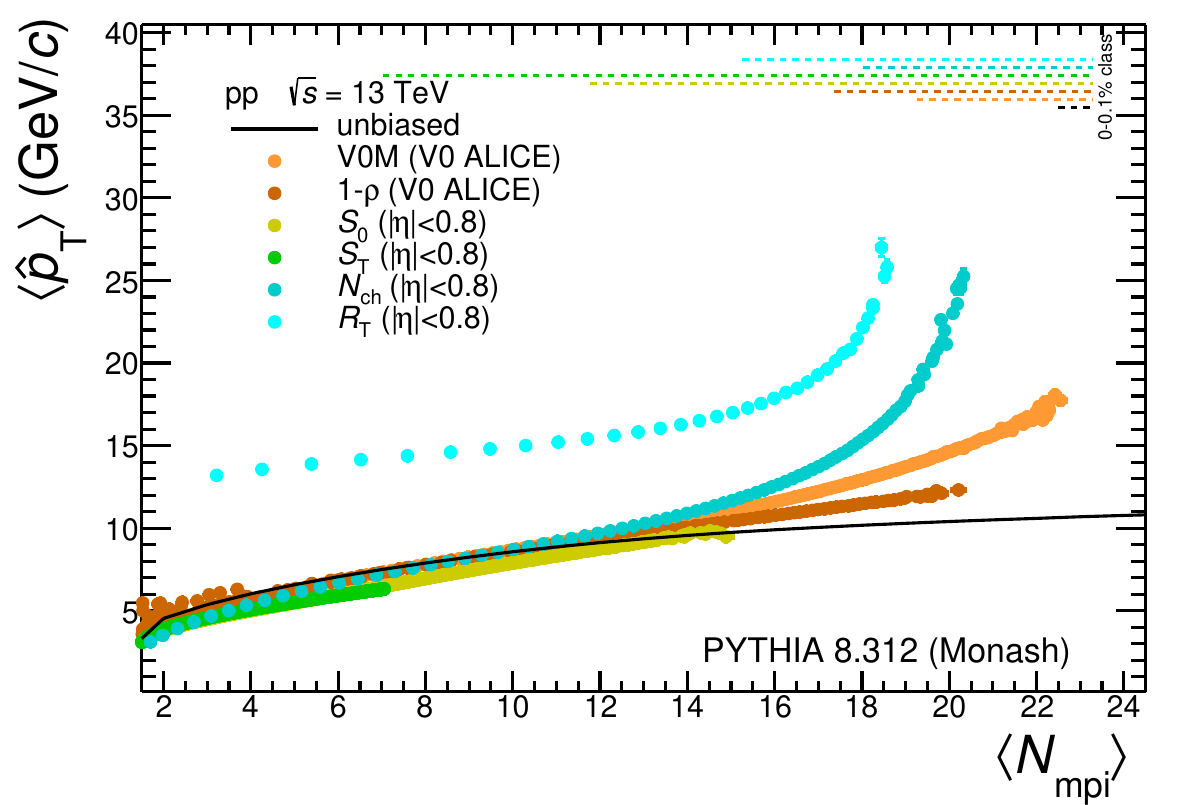}
\caption{Average $\hat{p}_{\rm T}$ (of the main parton-parton scattering) as a function of number of the multi-parton interactions in pp collisions at $\sqrt{s}=13$\,TeV. Results from different event activity classifiers are compared. The error bars represent the statistical uncertainty. The dashed lines indicate the \nmpi intervals covered by the top 0.1\% event classes.}
\label{fig:1}  
\end{figure}

The correlation between the average \pt of the main parton-parton scattering ($\hat{p}_{\rm T}$) and the number of multiparton interactions ($N_{\rm mpi}$) is shown in Fig.~\ref{fig:1} for the 6 billion minimum-bias \pp collisions. The average $\hat{p}_{\rm T}$ increases with $N_{\rm mpi}$ due to the impact-parameter dependence of $N_{\rm mpi}$. The more central the collision, the higher the probability of finding more than one parton-parton scattering within the same collision. Therefore, the probability of finding a hard scattering increases with $N_{\rm mpi}$. From now on, the selection based on $N_{\rm mpi}$ is named as ``unbias'' in the sense that since no information at hadron level is used, any effect can only be attributed to the implicit impact-parameter selection. In general, all event activity estimators, except \rt, show a similar behavior for $N_{\rm mpi}<14$. More precisely, for \rt the average $\hat{p}_{\rm T}$ is nearly twice that of the other estimators, and its lowest average \nmpi is 3. This is because \rt implies the selection of pp collisions with a leading-particle \pt above 5\,GeV/$c$, where the average particle density associated to the underlying event reaches a maximum~\cite{ALICE:2019mmy,ALICE:2022fnb}. For larger \nmpi ($>14$), the $\langle \hat{p}_{\rm T} \rangle$ deviates from the unbiased case. The largest deviation is seen for the multiplicity-based estimators ($N_{\rm ch}$ and V0M), whereas flattenicity gives a correlation that is closer to the unbiased case. Moreover, all event activity estimators, except spherocity and sphericity, exhibit a large dynamic range in \nmpi. Sphericity exhibits the smallest $N_{\rm mpi}$ reach ($<7$), followed by spherocity with a $N_{\rm mpi}$ reach of around 16. The other estimators show larger $N_{\rm mpi}$-reach values. 

\begin{figure}[!h]
\centering
\includegraphics[width=0.6\textwidth]{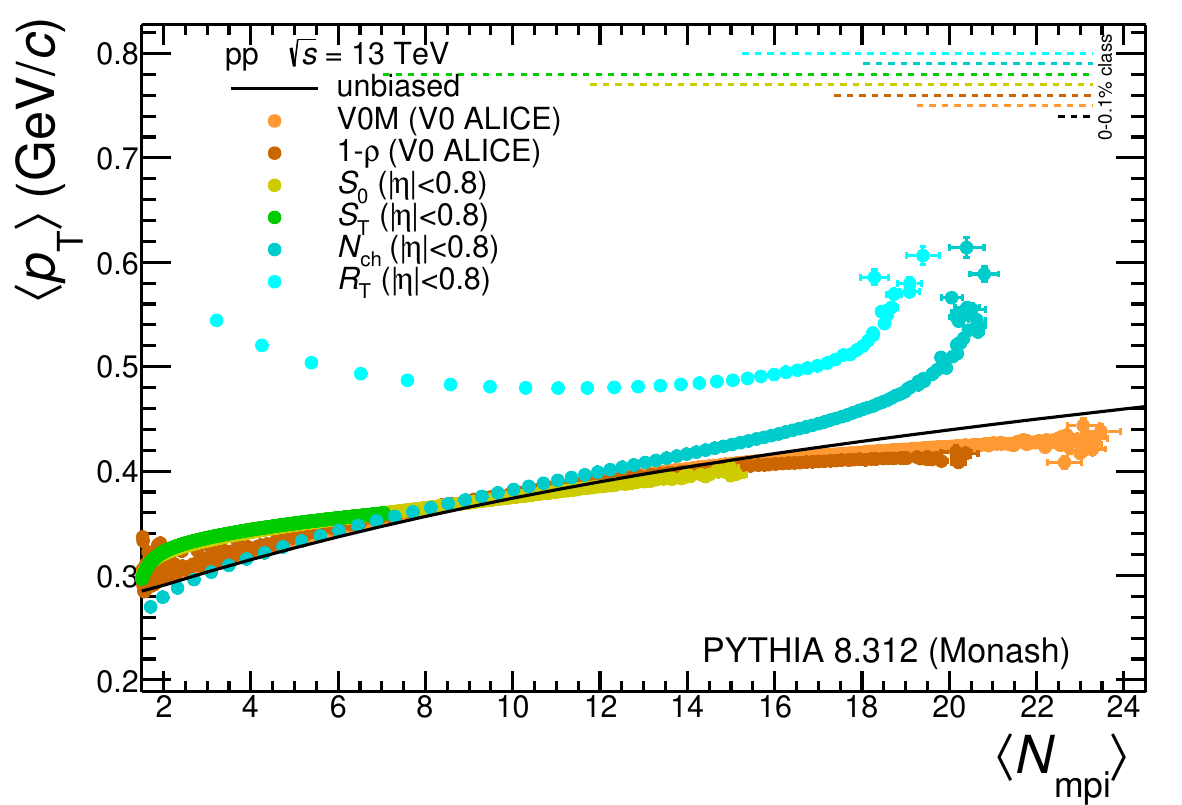}
 \caption{Average \pt of primary-charged particles as a function of number of multi-parton interactions in pp collisions at $\sqrt{s}=13$\,TeV. Results from different event activity classifiers are compared. The error bars represent the statistical uncertainty. The dashed lines indicate the \nmpi intervals covered by the top 0.1\% event classes.}
\label{fig:2}  
\end{figure}

\begin{figure}[!h]
\centering
\includegraphics[width=0.6\textwidth]{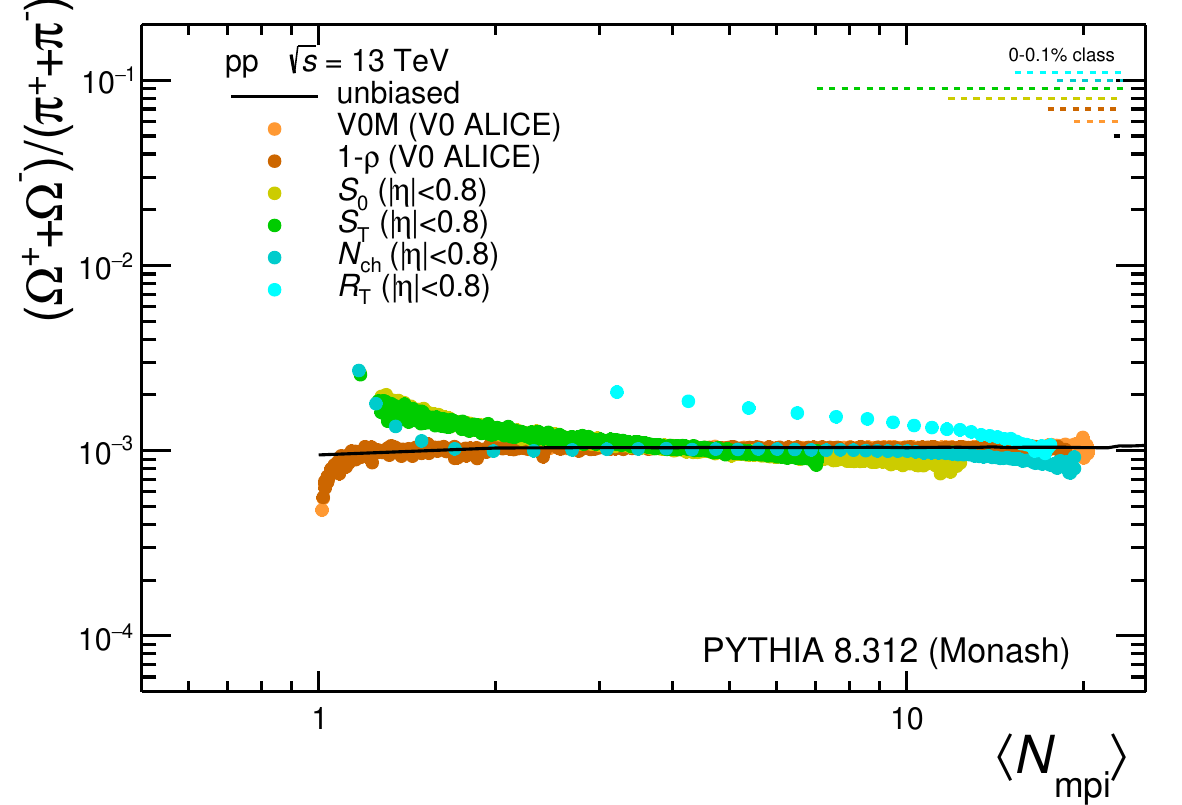}
\caption{Transverse momentum integrated omega baryon yield to charged pion yield as a function of \nmpi. Results for different event activity classifiers are shown. The error bars represent the statistical uncertainty. The dashed lines indicate the \nmpi intervals covered by the top 0.1\% event classes.}
\label{fig:3}  
\end{figure}

The impact at the hadron level is investigated by means of the average \pt and \pt-integrated particle ratios as a function of $N_{\rm mpi}$ for MB \pp collisions. For this study, only primary charged particles with $\pt > 0$ and within $|\eta|<0.8$ are considered. Figure~\ref{fig:2} shows that for the unbiased case, the average \pt of charged particles increases with increasing \nmpi. This general trend is well captured by the multiplicity estimators both at $|\eta|<0.8$ and at forward rapidity (V0M), as well as by flattenicity. However, at high \nmpi, the estimators at forward pseudorapidity exhibit a smaller average \pt than the expected one. Meanwhile, an opposite trend is observed for \rt and \nch, where an even higher average \pt is observed. In particular, \rt exhibits the highest deviation from all the event estimators in the whole range. Figure~\ref{fig:3} shows the \pt-integrated omega baryon yield normalized to the charged pion yield as a function of \nmpi. The results for \rt, $S_{\rm T}$, $S_{0}$ and $N_{\rm ch}$ give a yield ratio that is higher than the unbiased result (based on selection in $N_{\rm mpi}$) and actually exhibit an opposite trend, i.e., the yield ratio decreases with \nmpi increase.  Surprisingly, the yield ratios for V0M and flattenicity are closer to the unbiased case.

In order to investigate the biases for the most active pp collisions, i.e. $\approx0.1$\% of the cross section, selected with the different estimators; the thresholds indicated in Tab.~\ref{tab:0} were applied.

\begin{table}[!h]
\begin{center}
                \caption{ Fraction of the total cross section and the corresponding threshold value for the different event classifiers used in this work.
                 \label{tab:0}}
                \begin{indented}
                \item[]\begin{tabular}{lccc} 
                \br
                \textbf{Event classifier}  & \textbf{Symbol} &  \textbf{Percentile (\%)} & \textbf{Threshold} \\ 
                \mr
                 Multiplicity  &  $N_{\rm ch}$   & 0.11 & 59.5 \\
                 Sphericity      &  $S_{\rm T}$ &  0.11   & 0.983 \\
                 Spherocity      &  $S_{0}$ &  0.10  & 0.913 \\
                 Rel. trans. act. class.  &  $R_{\rm T}$   & 0.13 & 14.5 \\
                 Forward multiplicity     & V0M   & 0.10 &  129.5 \\
                 Flattenicity      &   1-$\rho_{\rm nch}$ & 0.11 & 0.910 \\
                 Multiparton interaction & $N_{\rm mpi}$ & 0.12 & 22.5 \\
                \br
                \end{tabular}
                \end{indented}
\end{center}
\end{table}

\section{Bias in the charged-to-neutral particles ratio}\label{sec:3}

\begin{figure}
\centering
\includegraphics[width=0.6\textwidth]{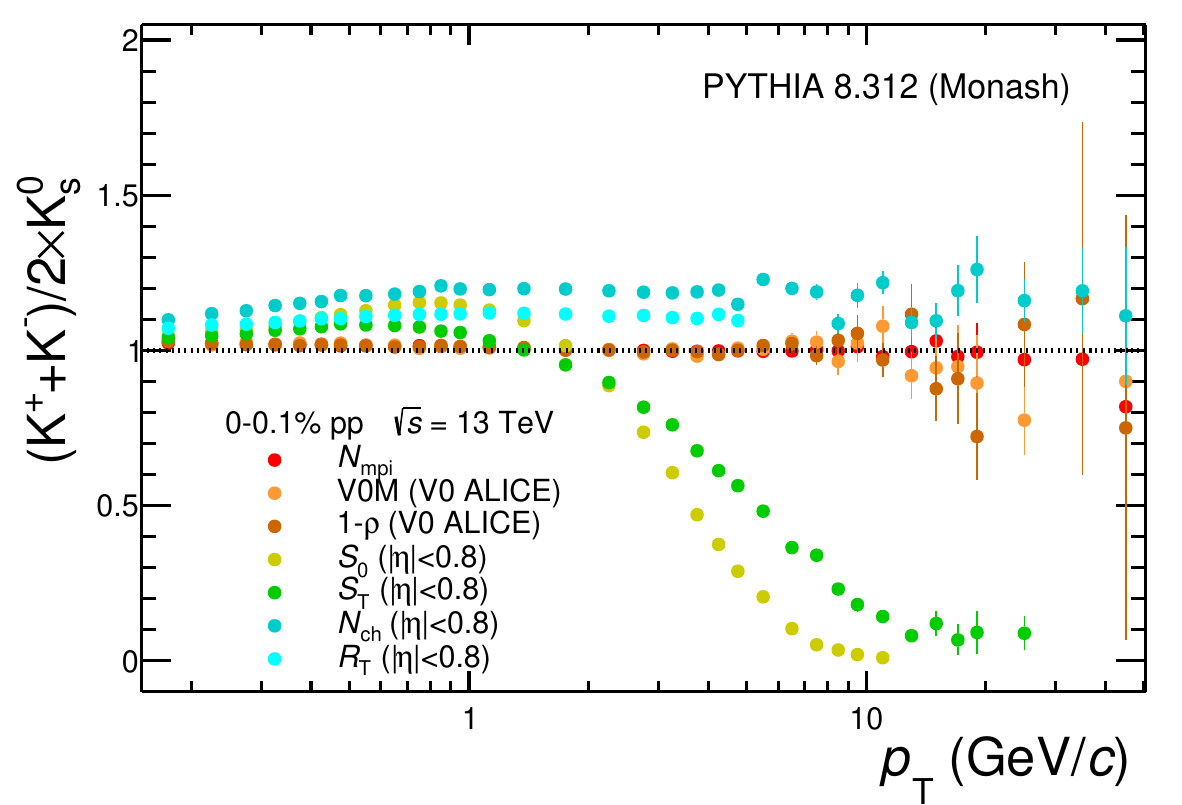}
\caption{Transverse momentum spectrum of charged kaons normalized to the neutral kaon \pt spectrum in the 0-0.1\% event classes. The error bars represent the statistical uncertainty.}
\label{fig:4}  
\end{figure}

One of the issues discussed in Ref.~\cite{ALICE:2016fzo} is the bias in the charged-to-neutral particle yields when measuring both the event activity and the observable of interest (e.g., the \pt spectrum) within the same pseudorapidity range. The observation of an increasing ratio with the increase in particle multiplicity at $|\eta|<0.8$ motivates the use of event classification based on the activity in the forward pseudorapidities covered by the ALICE V0 detector. Figure~\ref{fig:4} displays the ratio of charge-to-neutral kaons as a function of \pt for the 0-0.1\% event-activity class selected with different event activity variables. In pp collisions, this ratio should be close to unity due to isospin symmetry and, at high \pt, the fragmentation function of both ${\rm K}^\pm$ and ${\rm K}_{\rm s}^0$ are similar since both come from strange quark jets. Therefore, the event classifiers should ideally respect this behavior. However, both \rt and $N_{\rm ch}$ ($|\eta|<0.8$) give ratios that are above unity in the full \pt interval investigated in this work ($<50$\,GeV/$c$), reaching a discrepancy of up to 15\%. On the other hand, for transverse spherocity and sphericity the ratio is above unity for $\pt<1$\,GeV/$c$ and rapidly approaches to zero for larger \pt. This is because those event shapes do an implicit cut on high-\pt charged particles biasing the high-\pt neutral-to-charged particle ratio. The bias can be reduced by removing the \pt-dependence from the definition of event shapes~\cite{ALICE:2023bga,Prasad:2025yfj}. The aforementioned bias is also negligible if the event activity is calculated at forward rapidity like in the case of V0M and flattenicity estimators. For these cases, the ratios are fully consistent with unity in the full \pt interval. For completeness, Fig.~\ref{fig:4} also shows the ratio for the unbiased case (i.e. selection on \nmpi), which gives a ratio consistent with unity. It is worth mentioning that ALICE has reported the \pt-integrated ratio as a function of V0M multiplicity for pp collisions at 7\,TeV. Within uncertainties, such a ratio is fully consistent with unity for all the V0M event classes~\cite{ALICE:2018pal}.

\section{Bias towards hard physics}\label{sec:4}

\begin{figure}
\centering
\includegraphics[width=0.6\textwidth]{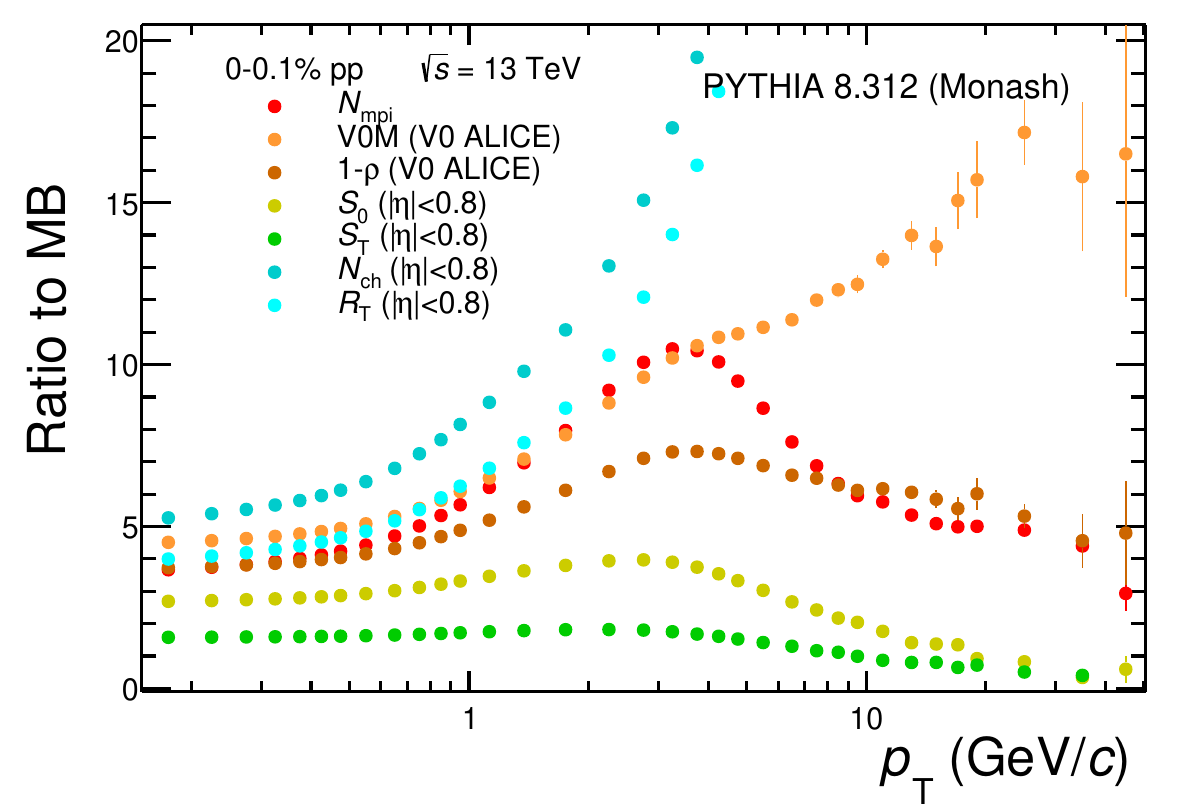}
\caption{Transverse momentum spectra for the 0-0.1\% event class normalized to the minimum bias \pt spectrum. Results for the event selection based on different event activity classifiers are presented. The error bars represent the statistical uncertainty.}
\label{fig:5}  
\end{figure}

The jet quenching searches in small systems require the event activity estimators to bias the jet fragmentation as little as possible. One can imagine that event shapes such as sphericity and spherocity will select mostly pp collisions with large underlying event relative to the jet signal. This is because, by construction, the highest event activity class will be sort of jet-free. Actually, Fig.~\ref{fig:5} shows that when plotting the transverse momentum spectrum for the 0-0.1\% event class, selected with event shapes, normalized to the spectrum in minimum-bias pp collisions, we see a suppression of the high-\pt particle yield. In contrast, for multiplicity estimators at mid-pseudorapidity ($N_{\rm ch}$ and \rt), the bias is the opposite: we select harder-than-average pp collisions.  The effect is slightly reduced when using the V0M multiplicity estimator. Therefore, the estimators based on charged particle multiplicity are biased towards harder-than-average pp collisions. However, flattenicity produces a ratio that is nearly flat at high \pt and consistent with the unbiased case (selection on \nmpi), suggesting that the event activity estimator does not bias the jet fragmentation relative to a minimum-bias pp collision. This feature is important for jet-quenching searches in small-collision systems.    

To study the correlation between event activity estimators and jet fragment multiplicity, FastJet version 3.4.3~\cite{Cacciari:2011ma} was implemented in the \py~8 simulations. A similar strategy to that reported in Ref.~\cite{ALICE:2023plt} was adopted. The 0-0.1\% event class \pp collisions were analyzed as follows. The leading particle within $|\eta|<0.8$ and the transverse momentum within 10-30\,GeV/$c$ was selected. The recoil jet was reconstructed using the anti-$k_{\rm T}$ algorithm implemented in FastJet. A radius of 0.4 was used and charged particles were considered in the analysis. The transverse momentum of the recoil jet is required to be greater than 25\,GeV/$c$ and this should be found in the opposite hemisphere relative to the leading particle ($|\varphi_{\rm jet}-\varphi_{\rm lead}|>\pi/2$). While in MB pp collisions, around 1.5 million events satisfy the selection criteria, only 18 and 182 events fulfill the requirements for the top 0.1\% sphericity and spherocity classes, respectively. However, for the top 0.1\% unbiased class (selection on \nmpi) around 12000 pp collisions are expected to satisfy the selection criteria. Interestingly, flattenicity is the estimator that yields a similar number for the top 0.1\% event class, around 11000 pp collisions. In contrast for the top 0.1\% V0M, \nch and \rt the numbers of recoil jets (same as number of pp collisions) are around 26000, 108000 and 242000, respectively.

\begin{figure}
\centering
\includegraphics[width=0.6\textwidth]{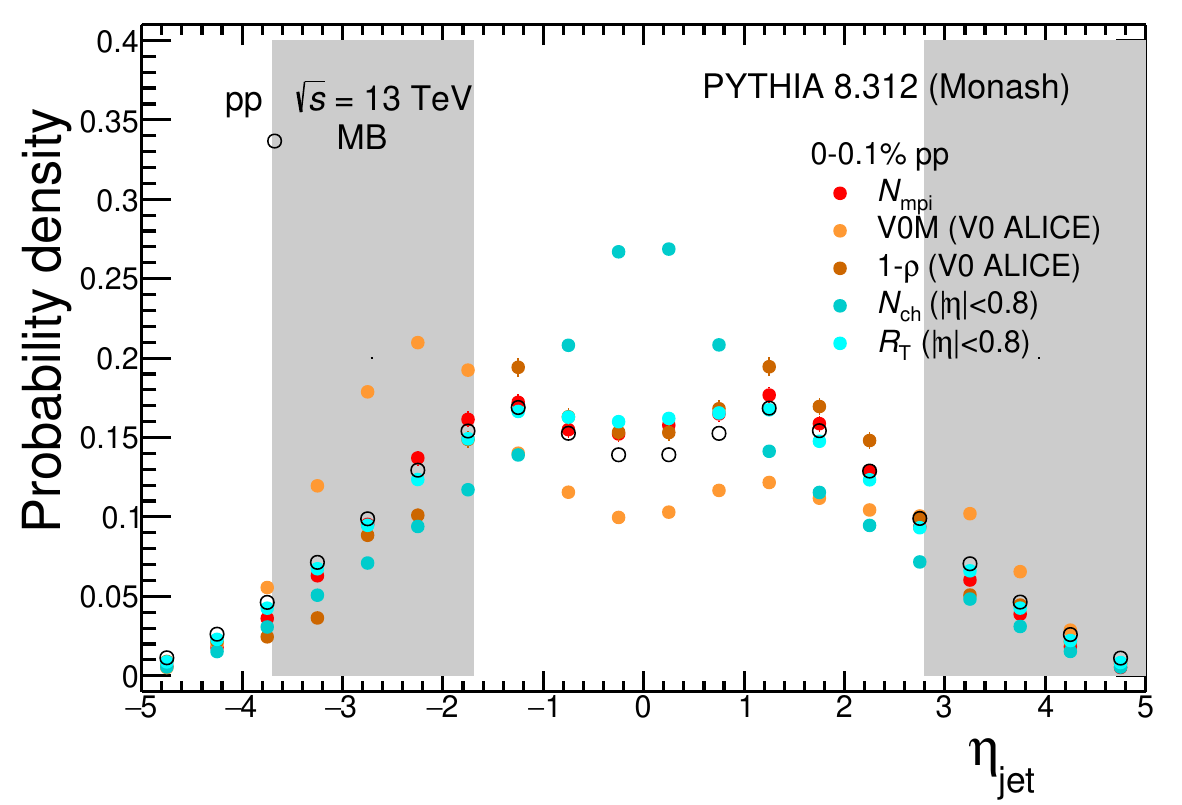}
\caption{Probability distribution of jets ($R = 0.4$) recoiling from a high-\pt hadron (within $10<p_{\rm T}^{\rm lead}<30$\,GeV/$c$ and $|\eta|<0.8$) as a function of $\eta_{\rm jet}$ for $p_{\rm T}^{\rm jet}>25$\,GeV/$c$, in pp collisions at $\sqrt{s}=13$\,TeV. Results for various event-activity estimators are shown in the 0-0.1\% event class. The shaded area represents the pseudorapidity regions covered by the ALICE V0 detector. The error bars represent the statistical uncertainty.}
\label{fig:6}  
\end{figure}

\begin{figure}
\centering
\includegraphics[width=0.6\textwidth]{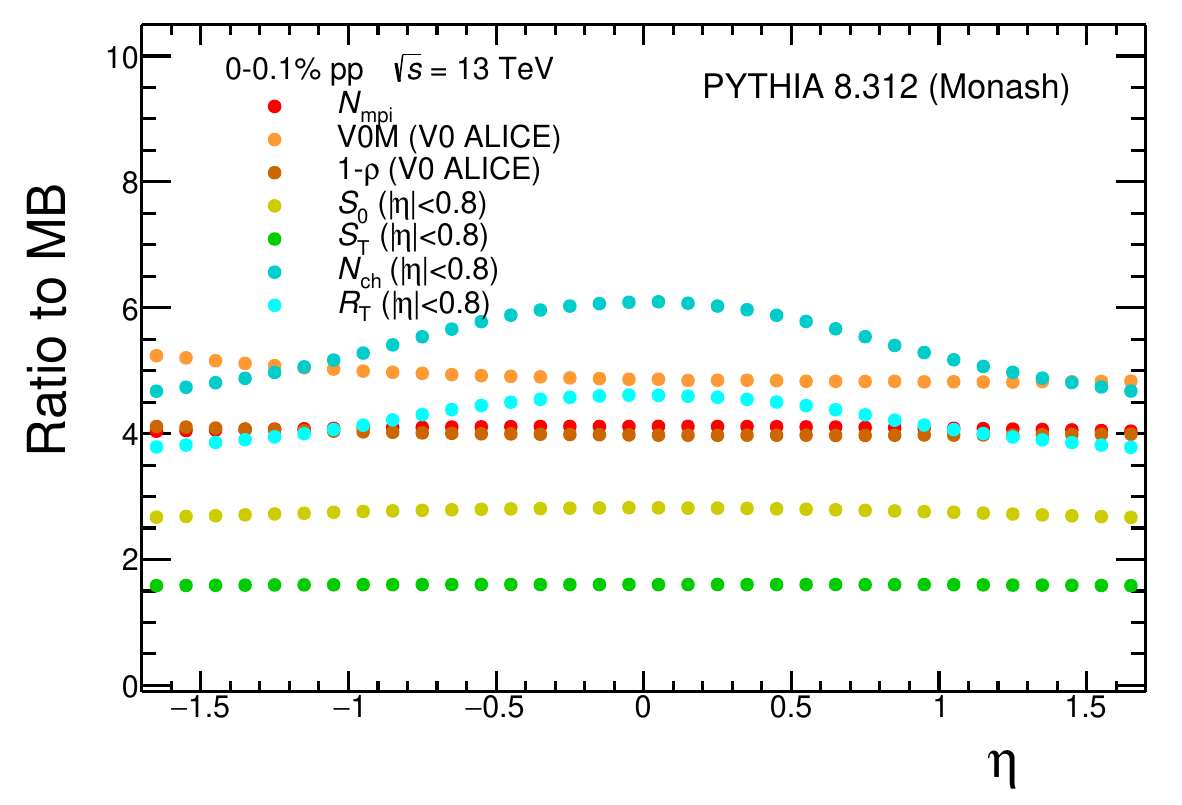}
\caption{The pseudorapidity distributions for the 0-0.1\% event class normalized to the minimum bias $\eta$ distribution. Results for different event activity estimators are compared. The error bars represent the statistical uncertainty.}
\label{fig:7}  
\end{figure}

The results of this study can be found in Fig. \ref{fig:6}, which shows the distribution of the jets recoiling from a high-\pt leading hadron as a function of jet pseudorapidity ($\eta_{\rm jet}$) for the different event-activity estimators. Sphericity and spherocity are missing in this figure since most of their corresponding recoil jets do not satisfy the selection criteria. While this distribution appears symmetric for the rest of the estimators, the V0M distribution presents an asymmetry, indicating a bias toward events with a recoil jet in either the V0C or the V0A region, shown as the shaded areas in the figure. Since V0C acceptance is narrower than that for V0A, there is a significant higher enhancement in the V0C region. Meanwhile, in the case of $R_{\rm T}$ and $N_{\rm ch}$ we can see an enhancement in the region where these estimators are calculated ($|\eta|<0.8$), indicating an autocorrelation or a bias towards coplanar jets. A notable feature of flattenicity is the absence of any of these biases and its resemblance to results from MB and unbiased (\nmpi selection) cases, demonstrating its potential to improve jet acoplanarity measurements~\cite{ALICE:2023plt}. A similar behavior is observed in Fig. \ref{fig:7} which shows the pseudorapidity distributions of primary charged particles for the 0-0.1\% for all the event classes normalized to the minimum-bias $\eta$ distribution. Although the distributions for $S_{T}$, $S_{0}$ and flattenicity exhibit a flat ratio, an increase in activity is observed for V0M, $R_{\rm T}$ and $N_{\rm ch}$ within the pseudorapidity intervals where these estimators are calculated. It should be noted that flattenicity produces a charged particle pseudorapidity density closer to that achieved in the unbiased case (selection on $N_{\rm mpi}$).

\section{Correlations between the different classifiers}\label{sec:5}

\begin{figure}
\centering
\includegraphics[width=0.6\textwidth]{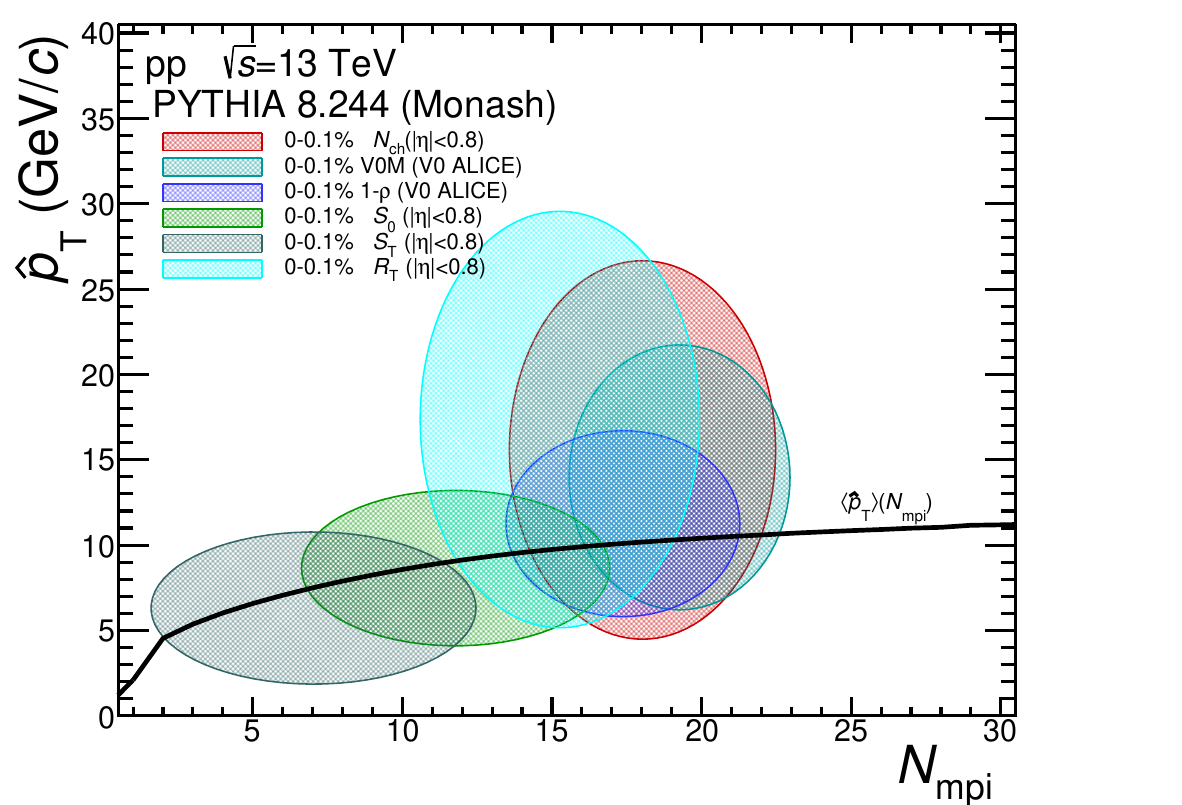}
\caption{Phase space in the $\hat{p}_{\rm T}$-\nmpi plane covered by the different 0-0.1\% event classes.}
\label{fig:8}  
\end{figure}

Figure~\ref{fig:8} shows the one-sigma area ($\hat{p}_{\rm T}$-$N_{\rm mpi}$) covered by each of the 0-0.1\% event classes selected with all the event estimators described in this paper. The estimators $S_{\rm T}$, $S_{0}$ and $1-\rho_{\rm nch}$ exhibit a narrower coverage in $\hat{p}_{\rm T}$ than the multiplicity-based estimators ($N_{\rm ch}$, V0M and $R_{\rm T}$). However, both $S_{\rm T}$ and $S_{0}$ show a larger coverage in MPI than flattenicity. Their corresponding mean $N_{\rm mpi}$ is around 7 and 12 for $S_{\rm T}$ and $S_{0}$, respectively; while for flattenicity a larger than 18 value is observed. More than 15 average $N_{\rm mpi}$ values are also seen for $N_{\rm ch}$, $R_{\rm T}$ and V0M, and their coverages in $\hat{p}_{\rm T}$ are very large. For $R_{\rm T}$, $N_{\rm ch}$ and V0M a one-sigma interval within the ranges $5<\hat{p}_{\rm T}<30$\,GeV/$c$, $4.5<\hat{p}_{\rm T}<26.5$\,GeV/$c$, and $6.5<\hat{p}_{\rm T}<22$\,GeV/$c$, respectively, can be observed. In contrast, flattenicity exhibits a narrower one-sigma coverage: $6<\hat{p}_{\rm T}<17$\,GeV/$c$.  In general, flattenicity gives the narrowest one-sigma covered region in $\hat{p}_{\rm T}$ and $N_{\rm mpi}$.  The overlap of most of the 0-0.1\% event classes ($R_{\rm T}$, $N_{\rm ch}$, V0M, flattenicity and $S_{0}$) is limited to a very narrow region around $N_{\rm mpi}\approx17$ and $\hat{p}_{\rm T}\approx10$\,GeV/$c$.    

Even though the same fraction of the cross section (0-0.1\%) is considered, the results in Fig.~\ref{fig:8} indicate that the selected events depend on the event classifier being used. Therefore, a direct comparison of the result obtained with different classifiers is not advisable. In the case of $R_{\rm T}$, $N_{\rm ch}$ and V0M, the broad dispersion in $\hat{p}_{\rm T}$ makes these classifiers more susceptible to harder-than-average pp collisions. This is not the case for $S_{\rm T}$, $S_{0}$, and flattenicity which exhibit a narrower dispersion in $\hat{p}_{\rm T}$ but provide a broader \nmpi coverage. Nonetheless, the \nmpi reach of both $S_{\rm T}$ and $S_{0}$ is limited when compared to flattenicity, which provides a decent \nmpi reach and a smaller deviation from the unbiased case.

Although flattenicity looks like the least biased event activity classifier, the event selection might still affect, for example, jet quenching searches. However, the analysis of both two LHC Runs 3 and 4 data as a function of flattenicity could reveal effects hidden by the standard multiplicity-based event activity classifiers.  In order to reduce the small biases, more information is required to measure the event activity. In this context, experiments such as ALICE 3 with unique particle identification capabilities of 8-10 units of pseudorapidity~\cite{ALICE:2022wwr,Dainese:2925455,Alfaro:2024sxc,MunozMendez:2025ttk}, will be crucial for the detailed study of small collision systems. 

\section{Summary}\label{sec:6}

This article reports the performance of different event-activity estimators, which have been reported by experiments at the LHC. The study has been conducted using pp collisions at $\sqrt{s}=13$\,TeV simulated with \py~8. The sample is equivalent to an integrated luminosity of around 70\,nb$^{-1}$. The list of estimators includes the multiplicity at midrapidity ($N_{\rm ch}$), forward multiplicity (V0M), sphericity, spherocity, flattenicty, and \rt.  The biases are reported for the top 0.1\% of the highest event activity. 

\begin{itemize}

\item The studies presented here suggest that the estimators with the highest dynamic range in \nmpi are flattenicity and those based on charged-particle multiplicity.
\item The bias on the neutral-to-charged particle yield is observed when selecting on \nch, \rt, sphericity and spherocity. It is worth mentioning that the ALICE Collaboration, recently proposed an approach to reduce this bias in observables like sphericity and spherocity. On the other hand, V0M and flattenicity are not affected by the bias.
\item For the 0-0.1\% event classes selected with most of the event-activity estimators ($R_{\rm T}$, $N_{\rm ch}$, V0M, $1-\rho_{\rm nch}$, and $S_{0}$), the overlap in the $\hat{p}_{\rm T}$-$N_{\rm mpi}$ space is very narrow. It is found around $N_{\rm mpi}\approx17$ and $\hat{p}_{\rm T}\approx10$, GeV /$c$. It is also interesting to note that the 0-0.1\% $S_{\rm T}$ class is outside such a common overlap region. This suggests that physics may differ from one event activity estimator to another.
\item For the 0-0.1\% event class,  FastJet was applied to reconstruct the recoil jet associated to leading particles with \pt within 10-30\,GeV/$c$. The pseudorapidity of the recoil jet was calculated for the different samples. By construction, pp collisions with high values of transverse spherocity and sphericity have little high-\pt particles at midrapidity and therefore, a negligible fraction of  recoil jets were reconstructed. For V0M, $N_{\rm ch}$ and $R_{\rm T}$, the $\eta_{\rm jet}$ distribution for the event class shows prominent structures in the pseudorapidity interval where the event activity estimator is calculated. This bias is significantly reduced or even absent in events triggered with flattenicity. This result encourages the use of flattenicity to search for jet quenching effects in small collision systems.  

\end{itemize}

\section*{\label{sec:conclusions}Acknowledgments}
Authors acknowledge the technical support from Luciano D\'iaz Gonz\'alez and Jes\'us Eduardo Murrieta Le\'on.  This work has been supported by DGAPA-UNAM PAPIIT No. IG100524 and PAPIME No. PE100124.

\section*{\label{sec:references}References}

\bibliographystyle{utphys}
\bibliography{cas-refs}

\end{document}